\documentclass[conference,letterpaper,10pt]{IEEEtran}
\usepackage{cite}
\usepackage{amsmath,amssymb,amsfonts}
\usepackage{algorithmic}
\usepackage{graphicx}
\usepackage{textcomp}
\usepackage{xcolor}

\newtheorem{theorem}{{{ \bf \textit{Theorem}}}}

\newtheorem{lemma}{{{ \bf \textit{Lemma}}}}
\newtheorem{corollary}{{{{ \bf \textit{Corollary}}}}}
\newtheorem{property}{{{\bf \textit{Property}}}}

\newtheorem{remark}{{{\bf \textit{Remark}}}}

\newtheorem{example}{{{\bf \textit{Example}}}}

\def\BibTeX{{\rm B\kern-.05em{\sc i\kern-.025em b}\kern-.08em T\kern-.1667em\lower.7ex\hbox{E}\kern-.125emX}}

\begin{document}

\title{On DNA Codes Over the Non-Chain Ring $\mathbb{Z}_4+u\mathbb{Z}_4+u^2\mathbb{Z}_4$ with $u^3=1$\\

}

\author{\IEEEauthorblockN{Shibsankar Das, Krishna Gopal Benerjee and Adrish Banerjee}
\IEEEauthorblockA{\textit{The Department of Electrical Engineering} \\
\textit{ Indian Institute of Technology Kanpur, UP, India}\\
Email: \textit{\{shibsankar, kgopal, adrish\}@iitk.ac.in}}
}

\maketitle


	\begin{abstract}
		In this paper, we present a novel design strategy of DNA codes with length $3n$ over the non-chain ring $R=\mathbb{Z}_4+u\mathbb{Z}_4+u^2\mathbb{Z}_4$ with $64$ elements and $u^3=1$, where $n$ denotes the length of a code over $R$. We first study and analyze a distance conserving map defined over the ring $R$ into the length-$3$ DNA sequences. Then, we derive some conditions on the generator matrix of a linear code over $R$, which leads to a DNA code with reversible, reversible-complement, homopolymer $2$-run-length, and $\frac{w}{3n}$-GC-content constraints for integer $w$ ($0\leq w\leq 3n$). Finally, we propose a new construction of DNA codes using Reed-Muller type generator matrices. This allows us to obtain DNA codes with reversible, reversible-complement, homopolymer $2$-run-length, and $\frac{2}{3}$-GC-content constraints.
	\end{abstract}	
	 	
	

\section{Introduction}
In recent years, the explosion of social networking has produced an ever-increasing amount of digital data \cite{2022Cao}. By $2025$, it is expected that $491$ exabytes (EB) of data per day will be accumulated around the world \cite{2025Seagate}. Due to this rapid data growth, storing such a massive amount of data with higher and efficient storage capacity becomes essential. Thus, it is important to investigate a new medium to store the data effectively. Owing to its higher storage capacity, lower energy consumption \cite{2012Church}, and outstanding storable time \cite{2016Bornholt}, the synthetic deoxyribonucleic acid (DNA) is an ideal medium to store a massive amount of digital data \cite{2015LimbachiyaIWSDA}. The DNA-based storage is extremely dense with a limit of $10^9  GB/mm^3$, and it has a long storable capability with a half-life of more than $500$ years \cite{2016Bornholt}.

The DNA strands can be considered as sequences of four nucleotides: A (Adenine), T (Thymine), G (Guanine), and C (Cytosine). The  Watson-Crick complement (or, simply complement) of DNA nucleotide is given by $\text{C}^c=$ G, $\text{G}^c=$ C, $\text{A}^c=$ T and $\text{T}^c=$ A. In DNA data storage, the digital data is translated into DNA sequences (constituted by four symbols A, T, G, and C). For example, two binary bits are mapped into a single nucleotide \cite{2017ErlichScience}. That is, $00 \rightarrow$ A, $10 \rightarrow$ G, $01 \rightarrow$ C and $11 \rightarrow$ T. To extract the original data, the DNA strings are sequenced and synthesized inversely into the binary data. 

In \cite{2018SongCOML}, it has been shown that two biochemical constraints, homopolymer run-length constraint and the fixed GC-content of DNA sequences, can be utilized to combat the sequencing and synthesizing errors in the retrieval process. The fixed GC-content constraint ensures that all DNA codewords enjoy a similar thermodynamic characteristic and the homopolymer run-length constraint limits the maximum number of consecutive repetitions of nucleotides in the encoded DNA \cite{2019WangCOML}\nocite{2018Immink}-\cite{2021Nguyen}.

In the oligonucleotide (i.e., single-stranded DNA) library, the non-specific DNA hybridization is a leading source of errors \cite{2008Chee}. To combat this unwanted DNA hybridization (or to avoid the errors in computation), the minimum Hamming distance between distinct DNA codewords, the reversible, and reversible-complement DNA codewords has been studied extensively in  \cite{2008Chee,2011Kim,2016Limbachiya,2018Dixita,2021KrishnaGopalCCDS,2021KrishnaGopalCOML,2022KrishnaGopalISIT}. Thus, it is more desirable that the oligonucleotide library should have a greater number of DNA codewords satisfying the maximum number of constraints for the given Hamming distance $d_H$.

In the design of DNA sequences from the ring structure, sequences are constructed by properly identifying a mapping from the ring into the set of DNA nucleotides \cite{2010Rocha}. The major advantage of using an algebraic approach over the computational methods is that it can produce an optimal DNA code set with a relatively longer sequence length \cite{2011Kim}. To date, there have been many algebraic construction methods for DNA sequences with certain constraints \cite{2006Abualrub,2013Oztas,2015Maheshanand,2016Bayram,2019Dinh,2020Liu,2020LiuAccess}. In \cite{2006Abualrub}, Abualrub \textit{et al.} investigated cyclic DNA sequences over a finite field $\mathbb{F}_4$. They studied DNA codes with reversible-complement constraint over $\mathbb{F}_4$. In \cite{2013Oztas}, Oztas and Siap studied reversible DNA codes over the finite field $\mathbb{F}_{16}$ with $16$ elements by introducing a new family of lifted polynomials. Later, in \cite{2015Maheshanand}, Srinivasulu and Bhaintwal investigated DNA codes with reversible-complement constraint over a finite chain ring $\mathbb{F}_{4}+u\mathbb{F}_4$ with $16$ elements and $u^2=0$. In \cite{2016Bayram}, Bayram \textit{et al.} studied DNA codes reversible constraint over a finite non-chain ring $\mathbb{F}_{4}+v\mathbb{F}_4$ with $v^2=v$. In \cite{2019Dinh}, Dinh \textit{et al.} proposed a construction of DNA codes with reversible constraint over a finite non-chain ring $\mathbb{Z}_{4}+u\mathbb{Z}_4$ with $16$ elements and $u^2=1$. In \cite{2020Liu}, Liu studied reversible DNA codes over a finite chain ring $\mathbb{F}_{4}[u]/\langle u^3\rangle$ with $64$ elements and $u^3=0$. They also investigated a construction method of reversible and reversible-complement DNA codes over a finite non-chain ring $\mathbb{Z}_{4}+v\mathbb{Z}_4$ with $16$ elements and $v^2=v$ in \cite{2020LiuAccess}. We aim to investigate DNA codes over $\mathbb{Z}_4+u\mathbb{Z}_4+u^2\mathbb{Z}_4$ with the maximum number of constraints and preferably better code-rate compared to existing work in the literature. 

In this paper, we study DNA sequences over a finite ring $R=\mathbb{Z}_4+u\mathbb{Z}_4+u^2\mathbb{Z}_4$ with $64$ elements and  $u^3=1$. We propose a design strategy for DNA codes with reversible, reversible-complement, $\frac{2}{3}$-GC-content, and homopolymer 2-run-length constraints over $R$. To be more specific, the main contributions of this paper are as follows:
\begin{itemize}
\item In this paper, we investigate a distance conserving bijective map $\psi$ between the ring $R$ and $\{\text{A, C, G, T}\}^3$. This mapping allows us to design DNA codes over $R$ with multiple constraints. 
\item Next, we study different properties of linear codes over $R$ with $u^3=1$ to construct DNA codes with reversible, reversible-complement, homopolymers $2$-run-length, and $\frac{2}{3}$-GC-content constraints.  
\item Finally, we propose a novel construction of new DNA codes satisfying multiple constraints of length $3\cdot 2^m$ over the finite non-chain ring $R$ by using Reed-Muller type generator matrices. 
\end{itemize}

The rest of the paper is organized as follows. In Section \ref{Pre:sec}, preliminaries and notations are presented. 
In Section \ref{Pro:Cons:sec}, the distance conserving map, Gau distance, and their properties are investigated.
Then, DNA codes with multiple constraints are constructed in Section \ref{Pro:Cons:DNA:sec}. 
Finally, a comparative analysis is drawn in Section \ref{Sec:Discussion}.

\section{Preliminaries}
\label{Pre:sec}
In this section, we study the structure of the ring $R=\mathbb{Z}_4+u\mathbb{Z}_4+u^2\mathbb{Z}_4$ with $u^3=1$. Then, we discuss some basic definitions and notations.

Let $R=\mathbb{Z}_4+u\mathbb{Z}_4+u^2\mathbb{Z}_4=\{a+bu+cu^2: a,b,c \in \mathbb{Z}_4\mbox{ and }u^3=1\}$. 
The ring $R$ is a commutative ring with identity and $64$ elements. 
An arbitrary element $x$ of the ring $R$ can be uniquely expressed by $x = a+bu+cu^2$, where $a,b,c \in \mathbb{Z}_4$. 
If we consider $u^3=1$, ideals of $R$ are not comparable. Thus, the ring $R$ is a non-chain ring. 
For the ring $R$, there are $24$ unit elements and $40$ zero divisors. 
The set of all $24$ unit elements is $\mathcal{U}$ = $\{2u+3u^2, 3u, 3u+2u^2, 1, 1+2u^2, 1+2u, 1+2u+2u^2, 2+u^2, 2+3u^2, 2+u, 2+u+2u^2, 2+2u+u^2, 2+2u+3u^2, 2+3u, 2+3u+2u^2, 3, 3+2u^2, 3+2u, 3+2u+2u^2, u^2, 3u^2, u, u+2u^2, 2u+u^2\}$.
Now, we define a sequence $\textbf{a}$ = $(a_1,a_2,\ldots,a_n)$ of length $n$ over $R$.
Further, for the sequence $\textbf{a}$, we define the  sequence $\textbf{a}^r$ = $(a_n,a_{n-1},\ldots,a_1)$.
Now, for any arbitrary matrix $G$ with $k$ rows $\textbf{g}_i$ ($i=1,2,\cdots,k$), $\langle G\rangle$ = $\left\{\sum_{i=1}^ka_i\cdot\textbf{g}_i:a_i\in R\right\}$
is the row sub-module of the matrix $G$ over $R$.
Again, for any given element $x\in R$, $\langle x\rangle$ = $\{a\cdot x:a\in R\}$. 
For any positive integer $n$, a set $C\subseteq R^n$ of size $M$ is called a code with parameter $(n,M,d_{\text{G}})$, where the minimum Gau distance $d_{\text{G}}$  will be defined in Section \ref{Gau distance}.
\begin{center}
\begin{figure}
\includegraphics[width=0.45\textwidth]{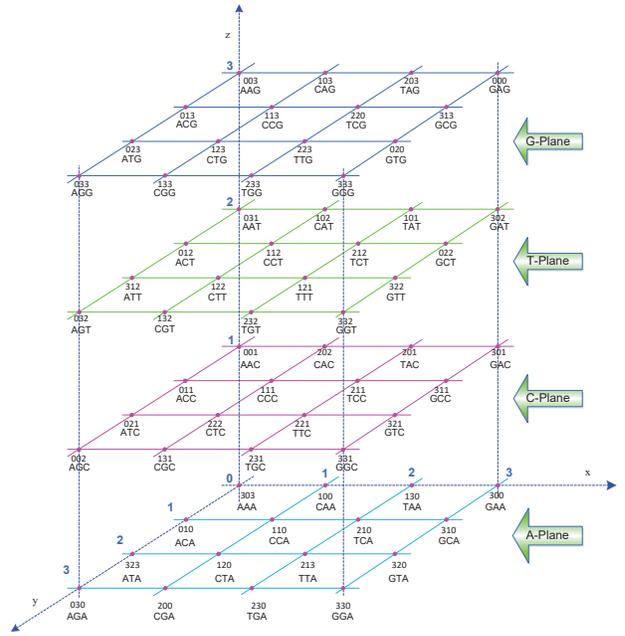}
\caption{Rearrangement of the elements of $R$ and the DNA sequences of length three in a three dimensional box. We denote this three dimensional box by $\Xi$. Here, an element $a+bu+cu^2\in R$ is simply denoted by `$abc$'.}
\label{fig:rearragement:3D:Box}
\end{figure}
\end{center}

A sequence $\textbf{c}$ = $(c_1,c_2,\ldots,c_n)$ of length $n$ over $\Sigma_{DNA}$ = $\{\text{A, C, T, G}\}$ is said to be a DNA sequence and also it can be represented by $c_1c_2\ldots c_n$. 
For any DNA sequence $\textbf{c}$ = $c_1c_2\ldots c_n$, the reverse DNA sequence is $\textbf{c}^r$ = $c_nc_{n-1}\ldots c_1$, and the reverse-complement DNA sequence is $\textbf{c}^{rc}$ = $c^c_nc^c_{n-1}\ldots c^c_1$, where $\text{C}^c=$ G, $\text{G}^c=$ C, $\text{A}^c=$ T and $\text{T}^c=$ A.
Also, the GC-content, $w_{GC}(\textbf{c})$, of the DNA sequence $\textbf{c}$ is the total number of G's and C's in the DNA sequence. 
In a DNA sequence, a homopolymer of run-length $\ell$ is the maximum number of consecutive positions $\ell$ where all the positions are $x$ for some $x\in\Sigma_{DNA}$.
For example, the DNA sequence $\text{CAAAGGT}$ of length $7$ has a homopolymer of maximum run-length $3$.
A set $C_{DNA}\subseteq \Sigma_{DNA}^n$ of size $M$ is called a DNA code with parameter $(n,M,d_H)$, where the minimum Hamming distance $d_H$ = $\min\{d_H(\textbf{x},\textbf{y}):\textbf{x}\neq\textbf{y}\mbox{ and }\textbf{x},\textbf{y}\in C_{DNA}\}$ and $d_H(\textbf{x},\textbf{y})$ is the Hamming distance between the DNA sequences $\textbf{x}$ and $\textbf{y}$.
The problem of designing DNA codes is to construct the desired DNA codes satisfying different constraints. 
There are four different constraints on $(n,M,d_H)$ DNA codes $C$ that are considered in this work:
\begin{itemize}
    \item \textbf{\textit{The Homopolymer $k$-Run-Length Constraint:}} In each codeword $\textbf{c} \in C$, if the maximum possible run-length is $k \ (\geq 1)$, then the DNA code satisfies the Homopolymer $k$-run-length constraint.
    \item \textbf{\textit{The $\frac{w}{n}$-GC-Content Constraint:}} For any DNA code with length $n$, if all DNA codewords have the same GC-content $w$, then the DNA code satisfies the $\frac{w}{n}$-GC-content constraint.
    \item \textbf{\textit{The Reversible Constraint:}} For any two codewords $\textbf{c}_1, \textbf{c}_2\in C$ such that $\textbf{c}_1^r\neq\textbf{c}_2$, the DNA code holds reversible constraint if $d_H(\textbf{c}_1^r,\textbf{c}_2)\geq d_H$.
    \item \textbf{\textit{The Reversible-Complement Constraint:}} For all $\textbf{c}_1, \textbf{c}_2\in C$ such that $\textbf{c}_1^{rc}\neq\textbf{c}_2$, the DNA code holds reversible-complement constraint if $d_H(\textbf{c}_1^{rc}, \textbf{c}_2)\geq d_H$. 
\end{itemize}

\section{DNA Codes over the Ring $R$}\label{Pro:Cons:sec}
In this section, we study and analyze a distance conserving map $\psi:R\rightarrow \Sigma_{\text{DNA}}^3$ (given in Table \ref{table:DNA:Seq:Ring:R}) with some key properties. 

\subsection{Distance Conserving Map}\label{Gau distance}
To construct the DNA codes over $R$, we first investigate a one-to-one correspondence between the length-3 DNA sequences and elements of $R$. We first rearrange all the $64$ elements of $R$ in a three-dimensional (3-D) box as given in Fig. \ref{fig:rearragement:3D:Box} to define the distance conserving map $\psi$. In Fig. \ref{fig:rearragement:3D:Box}, an element $a+bu+cu^2 \in R$ is denoted by `$abc$'. For example, $1+3u+2u^2$ is denoted by `$132$'.
In the 3-D box $\Xi$, the elements of $R$ and the DNA sequences of length three are placed such that the distance $d_H$ between two distinct DNA sequences in a single axis is 1, in two different axes is 2, and in three different axes is 3. 
This representation allows us to define the distance (i.e., Gau distance) between the elements of the ring $R$ so that the Hamming distance property is preserved. 
     
We consider that the positions of two elements $x$ and $y$ from $R$ are $(i,j,k)$ and $(i',j',k')$ in the 3-D box $\Xi$ with $0\leq i,j,k \leq 3$ and $0\leq i',j',k' \leq 3$. 
Then, the Gau distance $d_{\text{G}}:R\times R\rightarrow \mathbb{R}$ is defined by $d_{\text{G}}(x,y)$ = $\text{min}\{1,(i+3i')\hspace{-0.2cm}\mod 4\}+\text{min}\{1,(j+3j')\hspace{-0.2cm}\mod 4\}+\text{min}\{1,(k+3k')\hspace{-0.2cm}\mod 4\}$. 
For any integers $t$ and $t'$ ($0\leq t,t'\leq3$), we can see that $\min\{1,(t+3t')\hspace{-0.2cm}\mod4\}\geq0$, $\min\{1,(t+3t')\hspace{-0.2cm}\mod4\}$ = $0$ iff $t=t'$, and $\min\{1,(t+3t')\hspace{-0.2cm}\mod4\}$ = $\min\{1,(t'+3t)\hspace{-0.2cm}\mod4\}$.
Also, for any integers $t$, $t'$ and $t''$ ($0\leq t,t',t''\leq3$), we can observe that $\min\{1,(t+3t'')\hspace{-0.2cm}\mod4\}\leq\min\{1,(t+3t')\hspace{-0.2cm}\mod4\}+\min\{1,(t'+3t'')\hspace{-0.2cm}\mod4\}$. Therefore, the mapping $d_{\text{G}}$ satisfies the identity of indiscernible, symmetry, and triangle inequality. In other words, the mapping $d_{\text{G}}$ is a metric defined over $R$.

For any two vectors  $\textbf{x}=(x_1, x_2, \cdots, x_n)$ and $\textbf{y}=(y_1, y_2, \cdots, y_n)$ of equal length $n$ over $R$, the Gau distance between $\textbf{x}$ and $\textbf{y}$ is defined by $d_{\text{G}}(\textbf{x},\textbf{y})=\sum_{i=1}^{n}d_{\text{G}}(x_i,y_i)$. 
For any code $C$ over $R$, the minimum Gau distance $d_{\text{G}}$ is defined by $d_{\text{G}}=\text{min}\{d_{\text{G}}(\textbf{x},\textbf{y}): \textbf{x}, \textbf{y} \in C \ \text{with}\ \textbf{x}\neq \textbf{y} \}$.
For example, we consider $\textbf{x}=(2u, 1+3u^2,  2+u^2, 3) \in R^4$ and $\textbf{y}=(1+2u+u^2,3u, 2+2u, 3u+2u^2) \in R^4$. 
From the 3-D box $\Xi$, we have $d_{\text{G}}(2u,1+2u+u^2)=2$, $d_{\text{G}}(1+3u^2,3u)=3$, $d_{\text{G}}(2+u^2,2+2u)=2$ and $d_{\text{G}}(3,3u+2u^2)=3$. 
Thus, the Gau distance between $\textbf{x}$ and $\textbf{y}$ is $d_{\text{G}}(\textbf{x},\textbf{y})$ = $d_{\text{G}}(2u,1+2u+u^2)+d_{\text{G}}(1+3u^2,3u)+d_{\text{G}}(2+u^2,2+2u)$ $+d_{\text{G}}(3,3u+2u^2)$ = $2+3+2+3$ = $10$.

We now define the distance conserving map $\psi:R\rightarrow \Sigma_{\text{DNA}}^{3}$ such that, for any $x \in R$, 
\begin{align}
     &\psi(x)^c=\psi(x+(2+2u+2u^2)), \mbox{ and }  \label{comp:cons:eq}\\
     &\psi(x)^r=\psi(x^r),
\end{align}
where $x^r=c+bu+au^2$ for the element $x=a+bu+cu^2\in R$. By using the symmetry between the ring element and the DNA sequence of length $3$, we have developed the distance conserving mapping $\psi$ defined over the ring $R$ with $u^3=1$. We illustrate this bijective mapping $\psi:R\rightarrow \Sigma_{\text{DNA}}^{3}$ in Table \ref{table:DNA:Seq:Ring:R}. Note that this bijective mapping $\psi$ is not unique. For any $x\in R$, it is clear that $\psi^{-1}\left(\psi(x)\right)$, $\psi^{-1}\left(\psi(x)^c\right)$, $\psi^{-1}\left(\psi(x)^r\right)$, and $\psi^{-1}\left(\psi(x)^{rc}\right)$ are in the ring $R$.
For any $\textbf{x}=(x_1, x_2, \cdots, x_n)$ over $R$, we will define following:
\[
\begin{array}{l}
    \psi^{-1}\left(\psi(\textbf{x})\right) =  \left(\psi^{-1}\left(\psi(x_1)\right),\psi^{-1}\left(\psi(x_2)\right), \cdots \right. \\ 
    \hspace{5.5cm}\left.\cdots, \psi^{-1}\left(\psi(x_n)\right)\right), \\
    \psi^{-1}\left(\psi(\textbf{x})^r\right) =  \left(\psi^{-1}\left(\psi(x_n)^r\right),\psi^{-1}\left(\psi(x_{n-1})^r\right),\cdots\right. \\ \hspace{5.5cm}\left.\cdots,\psi^{-1}\left(\psi(x_1)^r\right)\right), \\
    \psi^{-1}\left(\psi(\textbf{x})^c\right) = \left(\psi^{-1}\left(\psi(x_1)^c\right), \psi^{-1}\left(\psi(x_2)^c\right), \cdots \right. \\
    \hspace{5cm}\left.\cdots,\psi^{-1}\left(\psi(x_n)^c\right)\right) \mbox{ and } \\
    \psi^{-1}\left(\psi(\textbf{x})^{rc}\right) = \left(\psi^{-1}\left(\psi(x_n)^{rc}\right), \psi^{-1}\left(\psi(x_{n-1})^{rc}\right), \cdots \right. \\
    \hspace{5.3cm}\left.\cdots,\psi^{-1}\left(\psi(x_1)^{rc}\right)\right).
\end{array}
\]
For any code $C$ of length $n$ over $R$, the DNA code of length $3n$ is given by  $\psi(C)=\{\psi(\textbf{x}): \mbox{ each } \textbf{x} \in C \}\subseteq \Sigma_{\text{DNA}}^{3n}$.
\begin{table*}[ht]
	\caption{Bijective map $\psi:R\rightarrow\Sigma_{DNA}^3$ for all  $x\in R$.}  \label{table:DNA:Seq:Ring:R}
      \centering 
		\begin{tabular}{|cl||cl||cl||cl||cl||cl}
			\hline
			$x$ & $\psi(x)$ & $x$ & $\psi(x)$ & $x$ & $\psi(x)$ & $x$ & $\psi(x)$ & $x$ & $\psi(x)$  & $x$ & \multicolumn{1}{c|}{$\psi(x)$} \\ 	\hline\hline
			$0$       & GAG & $3u+2u^2$ & AGT & $1+2u+u^2$  & TTT & $2+u$       & TCA & $2+3u^2$    & TAG & $3+3u+u^2$  & \multicolumn{1}{c|}{GGC} \\ \hline
		    $u$       & ACA & $3u^2$    & AAG & $1+3u+u^2$  & CGC & $2+2u$      & TCG & $2+u+3u^2$  & TTA & $3+2u^2$    & \multicolumn{1}{c|}{GAT} \\ \hline 
		    $2u$      & GTG & $u+3u^2$  & ACG & $1+2u^2$    & CAT & $2+3u$      & TGA & $2+2u+3u^2$ & TTG & $3+u+2u^2$  & \multicolumn{1}{c|}{ATT} \\ \hline 
		    $3u$      & AGA & $2u+3u^2$ & ATG & $1+u+2u^2$  & CCT & $2+u^2$     & TAC & $2+3u+3u^2$ & TGG & $3+2u+2u^2$ & \multicolumn{1}{c|}{GTT} \\ \hline 
		    $u^2$     & AAC & $3u+3u^2$ & AGG & $1+2u+2u^2$ & CTT & $2+u+u^2$   & TCC & $3$         & GAA & $3+3u+2u^2$ & \multicolumn{1}{c|}{GGT} \\ \hline 
		    $u+u^2$   & ACC & $1$       & CAA & $1+3u+2u^2$ & CGT & $2+2u+u^2$  & TTC & $3+u$       & GCA & $3+3u^2$    & \multicolumn{1}{c|}{AAA} \\ \hline 
		    $2u+u^2$  & ATC & $1+u$     & CCA & $1+3u^2$    & CAG & $2+3u+u^2$  & TGC & $3+2u$      & GTA & $3+u+3u^2$  & \multicolumn{1}{c|}{GCG} \\ \hline 
		    $3u+u^2$  & AAT & $1+2u$    & CTA & $1+u+3u^2$  & CCG & $2+2u^2$    & CAC & $3+3u$      & GGA & $3+2u+3u^2$ & \multicolumn{1}{c|}{ATA} \\ \hline 
		    $2u^2$    & AGC & $1+3u$    & TAA & $1+2u+3u^2$ & CTG & $2+u+2u^2$  & TCT & $3+u^2$     & GAC & $3+3u+3u^2$ & \multicolumn{1}{c|}{GGG} \\ \hline 
		    $u+2u^2$  & ACT & $1+u^2$   & TAT & $1+3u+3u^2$ & CGG & $2+2u+2u^2$ & CTC & $3+u+u^2$   & GCC &             &     \\ \cline{1-10} 
		    $2u+2u^2$ & GCT & $1+u+u^2$ & CCC & $2$         & CGA & $2+3u+2u^2$ & TGT & $3+2u+u^2$  & GTC &             &     \\ \cline{1-10} 
	\end{tabular}
\end{table*}

 \subsection{Properties of the Distance Conserving Map} 
In this section, we investigate the properties of distance conserving map and Gau distance. 
\begin{property}[Distance Conserving Property] 	\label{pro1:dis:preser:map}
 	The map $\psi$ is a distance conserving map from $(R^n,d_{\text{G}})$ to $\left(\Sigma_{\text{DNA}}^{3n}, d_{H}\right)$.
\end{property}

\begin{property}[Linearity Property w.r.t. Reverse]	\label{pro3:linearity::map}
For two length-$n$ vectors $\textbf{x}$ and $\textbf{y}$ in $R^n$, the map $\psi$ satisfies $\psi^{-1}\left(\psi(a\textbf{x}+b\textbf{y})^r\right)=a\psi^{-1}\left(\psi(\textbf{x})^r\right)+b\psi^{-1}\left(\psi(\textbf{y})^r\right)$.
\end{property}
From equation (\ref{comp:cons:eq}), we have the following result.
\begin{theorem}
\label{th:dis:close:complement}
For any given generator matrix $\mathcal{G}$ over $R$, the code $\psi(\left \langle \mathcal{G} \right \rangle)$ is closed under complement DNA sequences if ${\bf 2+2u+2u^2}\in \left \langle \mathcal{G} \right \rangle$, where ${\bf 2+2u+2u^2}$ denotes the vector of length $n$ with all $2+2u+2u^2$ element.
\end{theorem}

From \textit{Property \ref{pro3:linearity::map}}, we obtain \textit{Theorem \ref{th:dis:close:}}. 
Later, we will apply \textit{Theorem \ref{th:dis:close:}} to design DNA codes with reversible and reversible-complement constraints over $R$.
\begin{theorem}
	\label{th:dis:close:}
	The DNA code $\psi(\left \langle \mathcal{G} \right \rangle)$ is closed under reverse DNA sequence if $\textbf{g}^r=\psi^{-1}(\psi(\textbf{g})^r)\in \left \langle \mathcal{G} \right \rangle$ for each row $\textbf{g}$. 
\end{theorem}
\begin{IEEEproof}
For a matrix $\mathcal{G}$ with $k$ rows $\textbf{g}_1$, $\textbf{g}_2$, $\ldots$, $\textbf{g}_k$, let $\textbf{y} \in \left \langle \mathcal{G} \right \rangle$. 
To prove this result, we only need to show $\psi(\textbf{y})^r\in \psi(\left \langle \mathcal{G} \right \rangle)$. 
Since $\textbf{y} \in \left \langle \mathcal{G} \right \rangle$, there exist $a_1,a_2,\cdots, a_k \in R$ such that $\textbf{y}=\sum_{i=1}^{k}a_i\cdot\textbf{g}_i$. 
By using the property of the  operation $r$ on the element of $R$, we have $\textbf{y}^r=\sum_{i=1}^{k}a_i\cdot\textbf{g}_i^r$.
According to the given condition, each $\textbf{g}_i^r\in \left \langle \mathcal{G} \right \rangle$ for $i=1,2,\cdots, k$. 
Thus, $\textbf{y}^r=\sum_{i=1}^{k}a_i\cdot\textbf{g}_i^r \in \left \langle \mathcal{G} \right \rangle$. 
By applying \textit{Property \ref{pro3:linearity::map}}, we have $\textbf{y}^r\in \left \langle \mathcal{G} \right \rangle \Rightarrow \psi^{-1}\left(\psi(\textbf{y})^r\right) \in \left \langle \mathcal{G} \right \rangle \Rightarrow \psi(\textbf{y})^r\in \psi(\left \langle \mathcal{G} \right \rangle)$. 
\end{IEEEproof}

From \textit{Theorem \ref{th:dis:close:complement}} and \textit{Theorem \ref{th:dis:close:}}, we have \textit{Corollary \ref{coro:1:dna}}.
\begin{corollary}
\label{coro:1:dna}For a given $\mathcal{G}$ over $R$, the code $\psi(\left \langle \mathcal{G} \right \rangle)$ satisfies the reversible-complement constraint if $\psi(\left \langle  \mathcal{G} \right \rangle)$ is closed under both reverse and reverse-complement DNA sequences.
\end{corollary}
Consequently, we have the following result.
\begin{theorem}
\label{theorem:linear:code:DNA:code}
Let $C$ be a linear $(n,M,d_{\text{G}})$ code over $R$ and $\mathcal{G}$ be the associated generator matrix of $C$ such that the rows of $\mathcal{G}$ hold the conditions described in \textit{Theorem \ref{th:dis:close:complement}} and \textit{Theorem \ref{th:dis:close:}}, where $n, M$, and $d_{\text{G}}$ denote the length, size, and the minimum Gau distance of the code $C$. Then, $\psi(C)$ is a $(3n,M,d_H)$ DNA code with length $3n$, the size $M$, and the minimum Hamming distance $d_H$ = $d_{\text{G}}$. 
The DNA code $\psi(C)$ holds reversible and reversible-complement constraints.  
\end{theorem}
\begin{IEEEproof}
    From  \textit{Property \ref{pro1:dis:preser:map}}, \textit{Theorem \ref{th:dis:close:complement}} and \textit{Theorem \ref{th:dis:close:}}, the proof follows.
\end{IEEEproof}

For any $z$ from $ \{2, 2u, 2u^2\}$, we have the ideal generated by $z$ given by $\left\langle z \right\rangle=\{0, 2, 2u, 2u^2, 2+2u, 2u+2u^2, 2+2u^2, 2+2u+2u^2\}$. We have discussed the following cases.\newline 
 {\bf Case-I:} $y\in R\backslash\langle z\rangle$: In this case, the GC-content of $\psi(x)$ is not identical for all $x\in\langle y\rangle$. Also, all length-$3n$ DNA sequences defined over $\psi(\langle y\rangle)$ do not avoid the homopolymers. \newline
{\bf Case-II:} $y\in\{2+2u, 2u+2u^2, 2+2u^2, 2+2u+2u^2\}$: In this case, the GC-content of $\psi(x)$ is identical for all $x\in\langle y\rangle$, and also, all length-$3n$ DNA sequences defined over $\psi(\langle y\rangle)$ are free from the homopolymers of run-length more than 2. In this case, we can observe the following two sub-cases.\newline
    {\bf Sub-case-I:} $y\in\{2+2u, 2u+2u^2, 2+2u^2\}$: In this sub-case, the element $2+2u+2u^2\notin\langle y\rangle$, and therefore, we cannot construct DNA codes with reversible-complement constraint from linear code over $\langle y\rangle$ using \textit{Theorem \ref{th:dis:close:complement}}.\newline
    {\bf Sub-case-II:} $y=2+2u+2u^2$: In this sub-case, $\langle 2+2u+2u^2\rangle$ = $\{0,2+2u+2u^2\}$. The code-rate of any DNA code defined over $\langle 2+2u+2u^2\rangle$ is significantly small.
        
Thus, we are interested to study the properties of DNA codes defined over $\psi(\left\langle z \right\rangle)$ for $z \in \{2, 2u, 2u^2\}$. Now, we investigate homopolymer property for the obtained DNA sequences in \textit{Lemma \ref{Homo lemma}}.
\begin{lemma}
For any $z\in\{2, 2u, 2u^2\}$, consider a vector $\textbf{x}\in\left\langle z\right\rangle^n$.
The DNA sequence $\psi(\textbf{x})$ is free of all homopolymers of run-length more than $2$.
\label{Homo lemma}
\end{lemma}
\begin{IEEEproof}
For given any $z\in\{2, 2u, 2u^2\}$, the proof follows from the principle of mathematical induction on $n$. 
Base Case: For any $x\in\left\langle z\right\rangle$, $\psi(\langle x\rangle)$ = $\left\{GAG,CGA,GTG,AGC,TCG,\right.$ $\left.CAC,GCT,CTC\right\}$, and the DNA image $\psi(x)$ is free of all homopolymers of run-length more than $2$. 
Inductive Step: For a positive integer $m$, consider $(x_1,x_2,\ldots,x_m)\in\left\langle z\right\rangle^m$.
Then, assume the DNA image $\psi((x_1,x_2,\ldots,x_m))$ is independent from all homopolymers of run-length more than $2$.
Now, for $x_{m+1}\in\left\langle z\right\rangle$, the vector $(x_1,x_2,\ldots,x_m,x_{m+1})\in\left\langle z\right\rangle^{m+1}$.
In this case, one can observe the following.
\begin{enumerate}
    \item From the assumption, more than two consecutive positions are not identical in $\psi((x_1,x_2,\ldots,x_m))$.
    \item For any vector $(x_m,x_{m+1})\in\left\langle z\right\rangle^2$, more than two consecutive positions are not identical in $\psi((x_m,x_{m+1}))$. 
\end{enumerate}
And thus, more than two consecutive positions are not identical in the DNA image $\psi((x_1,x_2,\ldots,x_{m+1}))$.
So, from the definition of homopolymers, the DNA image $\psi((x_1,x_2,\ldots,x_{m+1}))$ is free from all homopolymers of run-length more than $2$.
Now, from mathematical induction on $n$, the result follows.
\end{IEEEproof}

We now investigate some important properties of the constructed DNA sequences as follows.
\begin{lemma}
For any $z\in\{2, 2u, 2u^2\}$, if $\textbf{x}\in\left \langle z\right \rangle^n$ then the GC-content of the $3n$-length DNA sequence $\psi(\textbf{x})$ is $2n$.
\label{GC lemma}
\end{lemma}
\begin{IEEEproof}
For any given $z\in\{2, 2u, 2u^2\}$, if $x\in\left\langle z\right\rangle$, then the GC-content of $\psi(x)$ is $w_{GC}(\psi(x))$ = $2$. Thus, for $\textbf{x}$ = $(x_1,x_2,\ldots,x_n)\in\left\langle z\right\rangle^n$ the GC-content of the DNA sequence $\psi(\textbf{x})$ is $w_{GC}(\textbf{x})$ = $\sum_{i=1}^nw_{GC}(\psi(x_i))$ = $2n$.
\end{IEEEproof}

\begin{theorem}
For any $z\in\{2, 2u, 2u^2\}$, we consider a linear code with the generator matrix $\mathcal{G}$ over  $\left \langle  z \right \rangle$ such that the rows of $\mathcal{G}$ hold the conditions described in \textit{Theorem \ref{th:dis:close:complement}} and \textit{Theorem \ref{th:dis:close:}}. Then, we have the following results: 
\begin{enumerate}
    \item The DNA code $\psi(\left\langle \mathcal{G} \right \rangle)$ satisfies reversible, and reversible-complement constraints.
	\item The GC-content of each DNA codeword of length $3n$ is $2n$.
	\item All DNA codewords are free from all homopolymers of run-length more than $2$.
\end{enumerate}
\label{GC Homopolymer}
\end{theorem}
\begin{IEEEproof}
From \textit{Lemma \ref{Homo lemma}}, all the DNA sequences in the DNA code $\psi(\left \langle  \mathcal{G} \right \rangle)$ are independent from homopolymers of run-length more than $2$.
Further, from \textit{Lemma \ref{GC lemma}}, all the DNA sequences in the DNA code $\psi(\left \langle  \mathcal{G} \right \rangle)$ satisfy $\frac{2}{3}$-GC-content constraint.
The result on reversible and reversible-complement properties follow from \textit{Theorem \ref{th:dis:close:complement}} and \textit{Theorem \ref{th:dis:close:}}.
\end{IEEEproof}

\section{Proposed Constructions of DNA Codes}\label{Pro:Cons:DNA:sec}
In this section, we present new constructions of DNA codes by using
Reed-Muller type codes over $R$.

For a given positive integer $m\  (\geq 1)$, the generator matrix $\mathcal{G}_{1,m}$ of the Reed-Muller type code $\mathcal{R}(1,m)$ of length $2^m$ is 
\begin{equation}
\label{gene:mat:R(1,m)}
\mathcal{G}_{1,i+1}=\begin{bmatrix}
\mathcal{G}_{1,i} & \mathcal{G}_{1,i} \\
\textit{\textbf{0}}_{2^i} & \textit{\textbf{z}}_{2^i}
\end{bmatrix}; \ 1\leq i\leq m-1,
\end{equation}where $\mathcal{G}_{1,1}=\begin{bmatrix}
z & z \\
0 & z
\end{bmatrix}$, $\textit{\textbf{0}}_{2^i}=\underbrace{[0\ 0\ \cdots \ 0]}_{2^i}$, $ \textit{\textbf{z}}_{2^i}=\underbrace{[z \ z \ \cdots \ z]}_{2^i}$ with $z \in R$. 

Clearly, the number of rows and number of columns of the generator matrix $\mathcal{G}_{1,m}$ are $(m+1)$ and $2^{m}$. 
\begin{remark}
\label{r RM code}
For any positive integer $m$ and $z\in R$, $\textbf{g}_i$ is the $i$-th row of $\mathcal{G}_{1,m}$ then $\textbf{g}_i^r$ = $\textbf{g}_1$ and $\textbf{g}_i^r$ = $\textbf{g}_1-\textbf{g}_i$ for $i=2,3,\ldots,m+1$.
Thus, $\textbf{g}_i^r\in\langle\mathcal{G}_{1,m}\rangle$ for $i=1,2,\ldots,m+1$.
\end{remark}
\begin{remark}
\label{c RM code}
For any positive integer $m$ and $z\in\mathcal{Z}\backslash\{2u+2u^2, 2+2u^2, 2+2u, 1+u+2u^2, 1+2u+u^2, 2+u+u^2, 2+3u+3u^2, 3+2u+3u^2, 3+3u+2u^2, u+3u^2, 3u+u^2, 1+3u^2, 1+3u, 3+u^2, 3+u\}$, we have $2+2u+2u^2\in\langle z
\rangle$. Thus, ${\bf 2+2u+2u^2}\in\langle\mathcal{G}_{1,m}\rangle$ as the vector with all $z$ is the row of $\mathcal{G}_{1,m}$, where $\mathcal{Z}$ is the collection of all zero divisors of $R$.
\end{remark}
Consequently, based on \textit{Theorem \ref{theorem:linear:code:DNA:code}}, we obtain \textit{Theorem \ref{unit RM code}}, \textit{Theorem \ref{theo:DNA:main}}, \textit{Theorem \ref{zero RM}} and \textit{Theorem \ref{theo2:DNA}}.
\begin{theorem}
For $z\in\mathcal{U}$, we consider the code $\mathcal{R}(1,m)$ over $R$. Then, the length, size and minimum Gau distance are $2^m$, $64^{m+1}$ and $d_{\text{G}}=2^{m-1}$.
\label{unit RM code}
\end{theorem}
\begin{IEEEproof}
For the $m$-th iteration, by using the principle of mathematical induction on $m$, the code length of $\mathcal{R}(1,m)$ is $2^{m}$. 
For $z\in\mathcal{U}$, if a matrix $\mathcal{G}_{1,i}$ over $R$ can be reduced to $[\textbf{I}_k\ \textbf{G}_{1,2}]$ then the size of $\langle\mathcal{G}_{1,i}\rangle$ is $64^k$, where $\textbf{G}_{1,2}$ is a matrix over $R$.
For any $x,y\in R$, the distance $d_{\text{G}}(x,y)\geq d_H(x,y)$. Note that $d_H(x,y)$ refers to the Hamming distance between the ring elements $x$ and $y$ (not to be confused with the Hamming distance between DNA sequences).
Thus, for the code $\mathcal{R}(1,m)$ with the minimum Gau distance $d_{\text{G}}$ and the minimum Hamming distance $2^{m-1}$, $d_{\text{G}}\geq 2^{m-1}$.
But, $\textit{\textbf{0}}_{2^{m}}$ and $\textit{\textbf{0}}_{2^{m-1}}||\textit{\textbf{3}}_{2^{m-1}}$ are the codewords of $\mathcal{R}(1,m)$ for any positive integer $m$, where $||$ denotes the concatenation of two sequences. Then, $d_{\text{G}}(\textit{\textbf{0}}_{2^m},\textit{\textbf{0}}_{2^{m-1}}||\textit{\textbf{3}}_{2^{m-1}})$ = $d_{\text{G}}(\textit{\textbf{0}}_{2^{m-1}},\textit{\textbf{3}}_{2^{m-1}})$ = $2^{m-1}$.
Thus, the minimum Gau distance is $d_{\text{G}}=2^{m-1}$.
\end{IEEEproof}

\begin{theorem}
\label{theo:DNA:main}
For the code $\mathcal{R}(1,m)$ over $R$, there exists a DNA code $\psi(\mathcal{R}(1,m))$, that satisfies both reversible and reversible-complement constraints with the length $3\cdot2^m$, size $64^{m+1}$ and minimum Hamming distance $2^{m-1}$.
\end{theorem}
\begin{IEEEproof}
The sequence length, size and minimum Hamming distance follow from \textit{Theorem \ref{theorem:linear:code:DNA:code}} and \textit{Theorem \ref{unit RM code}}. The reversible and reversible-complement constraints follow from \textit{Remark \ref{r RM code}}, \textit{Remark \ref{c RM code}}, \textit{Theorem \ref{th:dis:close:complement}} and \textit{Theorem \ref{th:dis:close:}}.
\end{IEEEproof}

\begin{example}\label{eg:1}
For $m=2$ and $z=3$, the $(n=12,M=262144,d_H=2)$ DNA code $\psi(\mathcal{R}(1,2))$ holds the reversible and reversible-complement constraints with the relative minimum Hamming distance $0.167$ and the code-rate $0.750$.
\end{example}
\begin{theorem}
For $z\in\{2,2u,2u^2\}$, 
the Reed-Muller type code $\mathcal{R}(1,m)$ over $R$, the length, size and minimum Gau distance are $2^m$, $8^{m+1}$ and $d_{\text{G}}=2^{m-1}$.
\label{zero RM}
\end{theorem}
\begin{IEEEproof}
The proof is similar to the proof of \textit{Theorem \ref{unit RM code}}.
Also, $\mathcal{G}_{1,m}$ can be reduced into $[2\textbf{I}_k\ 2\textbf{G}_{1,2}]$, since there exists $y\in R$ such that $x=2y$ for each $x\in\langle z\rangle$. Then, the size of $\langle\mathcal{G}_{1,m}\rangle$ is $8^{m+1}$, where $\textbf{G}_{1,2}$ is the matrix over $R$.
\end{IEEEproof}
\begin{theorem}
\label{theo2:DNA}
For $z\in\{2,2u,2u^2\}$ and the Reed-Muller type code $\mathcal{R}(1,m)$ over $R$, the DNA code $\psi(\mathcal{R}(1,m))$, satisfies Homopolymer $2$-run-length, $\frac{2}{3}$-GC-content, reversible, and reversible-complement constraints with length $3\cdot2^m$, size $8^{m+1}$ and minimum Hamming distance $2^{m-1}$.
\end{theorem}
\begin{IEEEproof}
The proof follows from Theorem \ref{GC Homopolymer}.
\end{IEEEproof}

\begin{example}\label{eg:2}
For $m=2$ and $z=2u$, the $(n=12,M=512,d_H=2)$ DNA code $\psi(\mathcal{R}(1,2))$ satisfies the Homopolymer $2$-run-length, $\frac{2}{3}$-GC-content, reversible and reversible-complement constraints with the relative minimum Hamming distance of $C$ is $0.167$ and the code-rate is $0.375$.
\end{example}

\begin{remark}
Note that the code-rate for above Reed-Muller type codes approaches zero as length approaches infinity. Thus, the code-rate of the constructed DNA codes vanishes for the large length. However, if we consider a linear code such that the code-rate does not vanish for the large length (e.g., Hamming type codes), then the code-rate for the generated DNA codes will also not vanish for the large length. Therefore, we can obtain DNA codes satisfying all four constraints such that the code-rate does not vanish for the large length.
\end{remark}
Our proposed design method is valid for DNA codes of length $3n$ (given in \textit{Theorem \ref{theorem:linear:code:DNA:code}}), where $n$ is the length of the code defined over $R$. In particular, when we consider the Reed-Muller type code of length $2^m$, we can construct DNA codes of length $3\cdot 2^m$ given in \textit{Theorem \ref{theo:DNA:main}}.


\begin{table}[ht]
\caption{Comparison with Existing Works}  \label{table:DNA:Seq:comparison}
\centering 
\begin{tabular}{|p{3cm}|p{1.4cm}|p{0.6cm}|l|l|l|}
\hline
Ring/Field and \newline DNA code   & Parameter \newline $(n,M,d_H)$ & Code 
\newline Rate & R & RC  & RLF \\ 
\hline
$\mathbb{Z}_{4}[u]/\langle u^3\mbox{-}1\rangle$; \textit{Example \ref{eg:1}} &  $(12,2^{18},2)$ & $0.750$ & Yes & Yes & -  \\ \hline 
$\mathbb{Z}_{4}[u]/\langle u^3\mbox{-}1\rangle$; \textit{Example \ref{eg:2}} &  $(12,512,2)$ & $0.375$ & Yes & Yes & Yes  \\ 
\hline
 $\mathbb{F}_4$; \cite[Table 1]{2006Abualrub} &  $(7,256,3)$ & $0.571$ & - & Yes  & - \\ 
\hline
$\mathbb{F}_{16}$; \cite[Example 4.6]{2013Oztas} &   $(18,4096,5)$ & $0.333$& - & Yes &  - \\ \hline
$\mathbb{Z}_{4}[u]/\langle u^2\mbox{-}(2\mbox{+}2u)\rangle$; \newline \cite[Table II]{2018Dixita} &  $(16,1024,8)$ & $0.313$ & Yes & Yes & -  \\ 
\hline
$\mathbb{F}_{4}[u]/\langle u^3\rangle$; \newline \cite[Example 2(3)]{2020Liu} &   $(15,256,10)$ & $0.267$ & Yes & Yes & -  \\ 
\hline
$\mathbb{F}_{4}[u]/\langle u^2\rangle$; \newline \cite[Example 3]{2015Maheshanand} &   $(12,64,4)$ & $0.250$ & - & Yes & -  \\ 
\hline
$\mathbb{F}_{16}$; \cite[Example 4.6]{2013Oztas} &   $(18,256,6)$ & $0.222$ & Yes & - & - \\ 
\hline
$\mathbb{F}_{4}[u]/\langle u^2\mbox{-}u\rangle$; \newline \cite[Example 5]{2016Bayram} &   $(20,256,8)$ & $0.200$ & Yes &  - & - \\ 
\hline
$\mathbb{Z}_{4}[u]/\langle u^2\mbox{-}u\rangle$; \newline \cite[Example 2]{2020LiuAccess} &  $(30,256,5)$ & $0.133$ & Yes & Yes &  - \\ 
\hline
\multicolumn{6}{p{8cm}}{Terms R, RC, and RLF refer to the $(n,M,d_H)$ DNA codes satisfying reversible, reversible-complement, and Homopolymer $2$-run-length constraints, respectively.}
\end{tabular}
\end{table}

\section{Discussions and Comparisons}\label{Sec:Discussion}
In this paper, for $z\in\mathcal{U}$, we have proposed algebraic construction for DNA codes $\psi(\mathcal{R}(1,m))$ that satisfies reversible, and reversible-complement constraints in \textit{Theorem \ref{theo:DNA:main}} with code-rate $(m+1)/2^m$. 
For $m=2$, an example of $(12,262144,2)$ reversible and reversible-complement DNA code is illustrated in \textit{Example \ref{eg:1}} with code-rate $0.750$.
Also, for $z\in\{2,2u,2u^2\}$, we have proposed algebraic construction for DNA codes $\psi(\mathcal{R}(1,m))$ that satisfies reversible, reversible-complement, $\frac{2}{3}$-GC-content and Homopolymer $2$-run-length constraints in \textit{Theorem \ref{theo2:DNA}} with code-rate $\frac{m+1}{3\cdot2^m}\log_48$. 
An example of such DNA code is illustrated in \textit{Example \ref{eg:1}} with parameter $(12,512,2)$ and code-rate $0.375$.
A comparison among properties of DNA codes is given in Table \ref{table:DNA:Seq:comparison}. 




	\end{document}